\documentclass[12pt]{iopart}

\usepackage{iopams}
\usepackage{setstack}
\usepackage{amscd}
\usepackage{hyperref}
\usepackage{epsf}

\newcommand\be{\begin{equation}}
\newcommand\ee{\end{equation}}
\newcommand\ba{\begin{eqnarray}}
\newcommand\ea{\end{eqnarray}}
\newcommand{\wteta}{{\widetilde{\theta}}}
\newcommand{\oteta}{{\overline{\theta}}}
\newcommand{\teta}{{\widetilde{\eta}}}
\newcommand{\eff}{{\rm eff}}
\newcommand{\narx}{\textit{Preprint} }
\newcommand{\arx}[1]{(\textit{Preprint} #1)}

\begin{document}

\rightline{\small Journal of Cosmology and Astroparticle Physics 10 (2005) 009  \hfill astro-ph/0411773}

\title[Non-Gaussianity in braneworld and tachyon inflation]{Non-Gaussianity in braneworld and\\ tachyon inflation}
\author{Gianluca Calcagni}
\address{Dipartimento di Fisica, Universit\`{a} di Parma, Parco Area delle Scienze 7/A, I-43100 Parma, Italy}
\address{Astronomy Centre, University of Sussex, Brighton BN1 9QH, UK}
\ead{g.calcagni@sussex.ac.uk}

\begin{abstract}
We calculate the bispectrum of single-field braneworld inflation, triggered by either an ordinary scalar field or a cosmological tachyon, by means of a gradient expansion of large-scale non-linear perturbations coupled to stochastic dynamics. The resulting effect is identical to that for single-field 4D standard inflation, the non-linearity parameter being proportional to the scalar spectral index in the limit of collapsing momentum. If the slow-roll approximation is assumed, braneworld and tachyon non-Gaussianities are subdominant with respect to the post-inflationary contribution. However, bulk physics may considerably strengthen the non-linear signatures. These features do not change significantly when considered in a non-commutative framework.
\end{abstract}


\maketitle


\section{Introduction}

According to the inflationary paradigm, early-Universe small quantum fluctuations of the scalar field driving a period of accelerated expansion are amplified to cosmological scales by the expansion itself. These perturbations then leave their imprint in the cosmic microwave background (CMB) as thermal anisotropies. Two main physical observables are generated by this mechanism, namely, the scalar spectrum, which is (the Fourier transform of) the two-point correlation function of scalar perturbations, and the bispectrum, coming from the three-point function \cite{GaM}--\cite{KSp}. For the curvature perturbation on comoving hypersurfaces $\cal R$, this reads 
\ba
\langle{\cal R}({\mathbf k}_1){\cal R}({\mathbf k}_2){\cal R}({\mathbf k}_3)\rangle &=& -(2\pi)^3\delta^{(3)}\left({\mathbf k}_1+{\mathbf k}_2+{\mathbf k}_3\right)\nonumber\\
&&\times\sum_{i<j} 2f_{\rm NL}({\mathbf k}_i, {\mathbf k}_j)\langle|{\cal R}_{\rm L}(k_i)|^2\rangle \langle|{\cal R}_{\rm L}(k_j)|^2\rangle,\label{fnl}
\ea
where $\langle|{\cal R}_{\rm L}(k)|^2\rangle$ is the power spectrum of the linear Gaussian part ${\cal R}_{\rm L}$ of the curvature perturbation with comoving wavenumber $k$, sum indices run from 1 to 3, and $f_{\rm NL}$ is the \emph{non-linearity parameter}\footnote{Our definition of the non-linearity parameter (sometimes dubbed $f_{\rm NL}^{\cal R}$ in literature) is $- 2\cdot 3/5$ that of \cite{GLMM,gan94} (there denoted $\Phi_3$), $- 3/5$ times that of \cite{KSp}, and $3/5$ times that of \cite{ABMR,mal03}. The factor of 3/5 comes from the definition of $f_{\rm NL}$ through the peculiar gravitational potential $\Psi_4$, which is $\Psi_4=-3{\cal R}_{\rm L}/5$ during matter domination.}.  If $f_{\rm NL}$ is momentum independent, one can write the gravitational potential in terms of ${\cal R}_{\rm L}$: in real space,
\be\label{nonlinear}
{\cal R}(\mathbf{x})={\cal R}_{\rm L}(\mathbf{x}) -f_{\rm NL}\left[{\cal R}^2_{\rm L}(\mathbf{x})- \left\langle {\cal R}^2_{\rm L}(\mathbf{x})\right\rangle \right]\,,
\ee
which gives equation (\ref{fnl}) with $f_{\rm NL}$ shifted outside the summation over the $\mathbf{k}_i$s. When the statistical distribution is Gaussian, $f_{\rm NL}=0$, the three-point function vanishes. In terms of the CMB temperature fluctuation $\Delta T(\mathbf{\hat{e}})/T$, measured along the direction $\mathbf{\hat{e}}$, the limit of the bispectrum at zero angular separation is the skewness, ${\cal S}_3(\mathbf{\hat{e}})\equiv \langle(\Delta T/T)^3\rangle$. For practical purposes, this is a less sensitive probe for non-Gaussianity than the bispectrum \cite{KSp}.

It is common to restrict the discussion of cosmological perturbations to the power spectrum, implicitly assuming dealing with all the relevant informations that can be extrapolated from the sky. In many respects, the measured two-point correlation function is able, all by itself, to both describe the microwave sky in great detail and place observational constraints on the features of early-Universe models such as inflation.

However, the experiments of the last generation have radically changed the general attitude towards cosmology and made possible what is now recognized as a `precision era'. The physical scenarios explaining the large-scale structure of the Universe can be refined by more and more accurate observational inspections. Therefore it is natural to consider the bispectrum, too, and ask what signatures of non-Gaussianity we might expect from a given theoretical model (see \cite{BKMR} for a comprehensive review).

A non-Gaussian spectrum can arise according to a number of different mechanisms: just to mention some, late-time non-linear evolution of cosmic structures \cite{MVH}, multi-field inflation or inflation with scalar spectators \cite{AGW}--\cite{EV}, the curvaton scenario \cite{LyW}--\cite{curv2}, non-vacuum inflation \cite{LPS}--\cite{GMS1},  higher-dimension operators in the inflaton Lagrangian \cite{cre03}, the inflaton with Dirac-Born-Infeld (DBI) Lagrangian \cite{AST}\footnote[1]{This model is quite distinct from the DBI cosmological tachyon. While the former predicts a strong non-Gaussian signature, the latter is more similar to standard inflation.}, the ghost inflaton \cite{ACMZ}, and, more commonly, a self-interacting inflaton \cite{OLM}--\cite{YV3}.

In four dimensions, it turns out that the inflationary contribution to the non-linear parameter is, for a single ordinary scalar field \cite{ABMR,mal03}, 
\be\label{fns}
4f_{\rm NL} \approx n_{\rm s}-1\,,
\ee
where $n_{\rm s}$ is the scalar spectral index 
\be
n_{\rm s}-1 \equiv 3+\frac{\rmd \ln \langle|{\cal R}_{\rm L}(k)|^2\rangle}{\rmd \ln k}\,.\label{ndef}
\ee
The scalar index can be written as a function of the slow-roll (SR) parameters describing the inflationary dynamics \cite{LL,SL}. At lowest SR order, that is in the quasi-de Sitter regime typical of inflationary models, the spectral index and the non-linearity parameter are first order in the SR parameters, which we denote symbolically as $f_{\rm NL}=\Or(\epsilon)$.

It would be interesting to see how this result is modified for braneworld cosmologies and tachyon-driven inflation.
In braneworld scenarios (e.g., \cite{mar03} and references therein) the visible universe and its matter content are confined into a (3+1)-dimensional Friedmann--Robertson--Walker (FRW) variety (a brane) embedded in a larger non-compact spacetime (called bulk). In the high-energy limit, the effective Friedmann equation experienced by a brane observer is, in suitable energy units,
\be \label{FRW}
H^2=\rho^q\,,
\ee
where $H$ is the Hubble parameter, $\rho$ is the energy density of the matter content (assumed to be a perfect fluid), and $q$ is a constant. The Randall--Sundrum (RS) and Gauss--Bonnet (GB) braneworlds display a more complicated evolution, but in particular time intervals or energy `patches' equation (\ref{FRW}) is a good approximation, with $q=2$ in the high-energy RS case and $q=2/3$ in the high-energy GB case. Also, tachyon cosmology can be implemented in the same formalism with little extra effort.

Intuitively, we do not expect a dramatic quantitative change in the non-Gaussian effect since the patch formalism does not interfere with SR expansions except for as regards the value of the coefficients in front of the SR parameters themselves \cite{cal3,cal5}. We have verified this guess explicitly. 

A preliminary confirmation comes from the stochastic approach of Gangui \textit{et al} \cite{GLMM}. This approach permits one to estimate the order of magnitude of the effect by just considering second-order fluctuations of a self-interacting inflaton field and no gravitational fluctuations. This might seem too crude an approximation, since one should go up to second order in perturbation theory in order to fully take gravitational back-reaction into account and treat the bispectrum consistently. Surprisingly, the inflaton perturbation really encodes the main feature of the model apart from the resulting incorrect combination of SR parameters.

Stochastic inflation is an approximated method according to which the solutions of the equation of motion for the scalar perturbation in the long wavelength limit $k \ll aH$ are connected to those in the $k \gg aH$ limit at the Hubble horizon \cite{sta86}--\cite{YV1}. Let $\psi$ denote the scalar field on the brane. This (or its gauge-invariant version) is separated into a `classical' or coarse-grained contribution $\psi^{(c)}$, encoding all the modes larger than the Hubble horizon, and a quantum or fine-grained part $\psi^{(q)}$ taking into account the in-horizon modes. Therefore, the classical part is the average of the scalar field on a comoving volume with radius $\sim (aH)^{-1}$. With this decomposition, the equation of motion for $\psi$ becomes a Langevin equation with a stochastic noise source generated by the fine-grained contribution of the quantum fluctuations (see \cite{MMR} for an extension to stochastic inflation with coloured noise).

Actually, we have performed the calculation of \cite{GLMM} with the general FRW equation (\ref{FRW}) and the potential SR tower \cite{cal3} both for the scalar field and the tachyon. So, $f_{\rm NL} = \Or(\epsilon)$ for a generic braneworld filled with a scalar field. As we shall see, the key assumption giving rise to this behaviour is the validity of the standard continuity equation; see equation (\ref{Nconti}) below. In fact, the presence of a brane-bulk flow would introduce genuinely novel features and considerably complicate the set of dynamical equations. Also, in the presence of extra dimensions the gravitational contribution may lead to a non-trivial behaviour of second-order perturbations, since to this order the interplay between extra-horizon scales and small scales may become quite delicate. This would impose a more rigorous treatment and a full second-order calculation in order to carefully evaluating the non-Gaussianity produced during inflation, e.g., one like that performed in \cite{ABMR} for the four-dimensional case.

A powerful approach in this sense is the space-gradient formalism of \cite{RiS1}--\cite{LyMS}, a development of the separate universe method \cite{WMLL} (see also the earlier works \cite{EB}--\cite{SaT}). We shall follow the notation of the cited papers closely, and skip details that can be found there unaltered by brane or tachyon physics. 

In this paper we achieve three goals. The first is to demonstrate the validity of the Mukhanov equation for the braneworld at large scales (section \ref{nlmukeq}); this equation allows us to calculate the power spectrum of the gauge-invariant scalar field fluctuation $u\propto a\delta\psi$ through a SR expansion \cite{muk85}--\cite{MFB}, and was long suspected to be valid even in the braneworld case. Not only does the method that we shall adopt give independent support to other proofs (e.g., \cite{RL}--\cite{der04} for the RS brane), but also it is valid beyond the linear order in perturbation theory. Also, our equation holds for a general patch and not only for the Randall--Sundrum case. The second outcome is the bispectrum of perturbations generated by either a cosmological tachyon or scenarios, such as high-energy braneworlds, with a modified Friedmann equation (section \ref{bispectra}); so far, only in \cite{cal3,der04} have some (very preliminary) remarks been made on the subject. Finally, in section \ref{ncbisp} these results are extended to the case of a maximally symmetric non-commutative spacetime \cite{BH}. Non-commutative
cosmology has recently drawn a lot of attention due to its possible effects on the shape of the power spectrum (for a list of references, see \cite{cal4,CT}); here we shall inspect for the first time the bispectrum predicted by this class of models. Conclusions are given in section \ref{concl}.


\section{Set-up}\label{ngsetup}

To begin with, it is convenient to work in the induced metric \cite{ArDM}
\be
\rmd s_4^2\Big|_{\rm brane} = N^2(t,\mathbf{x})\rmd t^2-a^2(t,\mathbf{x})\rmd x_i\rmd x^i\,,
\ee
where $N(t,\mathbf{x})$ is the lapse function ($N=1$ in synchronous gauge), $a(t,\mathbf{x})$ is a locally defined scale factor, and $\{x_1,x_2,x_3\}$ are space comoving coordinates on the brane. The Hubble parameter reads $H=\dot{a}/(Na)$; dots (primes) will denote derivatives with respect to $t$ ($\psi$). Also, we define $\Pi\equiv\dot{\psi}/N$ and the parameter $\theta \equiv 2(1-q^{-1})$. The first three SR parameters, which will be all we need, are
\ba
\epsilon &\equiv& -\frac{\dot{H}}{NH^2}=\frac{3q}{2}H^{\wteta} \frac{\Pi^2}{H^2}\,,\label{Npsepsi}\\
\eta     &\equiv&  -\frac{\dot{\Pi}}{NH\Pi}\,,\label{NHeta}\\
\xi^2    &\equiv& \frac{1}{NH^2} \left(\frac{\dot{\Pi}}{N\Pi}\right)^\cdot\,,
\ea
where $\wteta=\theta$ for the ordinary scalar and $\wteta=2$ for the tachyon. The SR evolution equations are
\ba
\dot{\epsilon} &=& NH\epsilon \left[\left(2-\wteta\right)\epsilon-2\eta\right]\,,\label{evo1}\\
\dot{\eta}     &=& NH\left(\epsilon\eta-\xi^2\right)\,.\label{evo2}
\ea
In the slow-roll approximation, $\epsilon,|\eta|,\xi^2\ll 1$, the inflaton potential energy $V(\psi)$ dominates other the kinetic term and the accelerated expansion is quasi-de Sitter ($H,\epsilon,\eta\approx$ constant). Sometimes a subscript in the SR symbols will specify the scalar model. For an ordinary scalar field $\phi$, the equation of motion at large scales reads ($\Pi=\dot{\phi}/N$)
\be\label{Piphi}
\dot{\Pi}+3NH\Pi+NV'(\phi)=0\,,
\ee
while for the DBI tachyon $T$ one has ($\Pi=\dot{T}/N$)
\be\label{PiT}
\frac{\dot{\Pi}}{1-\Pi^2}+3NH\Pi+N\ln V'(T)=0\,.
\ee
These are nothing but the continuity equation
\be \label{Nconti}
\dot{\rho}+3NH\rho (1+w) = 0\,,
\ee
where the barotropic index $w(p,\rho)$ relates the pressure and energy density of the perfect fluid filling the brane, $p=w\rho$.

In the separate universe approach, the physical quantities such as $H(t,{\mathbf x})$, the scalar field $\psi(t,{\mathbf x})$, cosmological perturbations and observables are defined on an inhomogeneous background and evolve separately through the dynamical equations at each point once the initial conditions have been specified. 
Then we can transform time derivatives into spatial gradients like
\be\label{spgrad}
\frac{\partial_i H}{H}=-\epsilon\frac{H}{\Pi}\partial_i\psi\,,\qquad \partial_i\epsilon=2\epsilon\left[\left(\oteta-\frac{\theta}{2}\right)\epsilon-\eta\right]\frac{H}{\Pi}\partial_i\psi\,,
\ee
where the index $i$ runs from 1 to 3, $\oteta=1$ for the ordinary scalar field, and $\oteta=\theta/2$ for the tachyon.

At large scales, second-order gradient terms can be neglected and the gradient generalization of the Bardeen potential \cite{BaST} is conserved \cite{RiS1}:
\ba
\zeta_i &\equiv& X_i-NH\frac{\partial_i\rho}{\dot{\rho}}\,,\\
\dot{\zeta}_i &=& 0\,,\label{conseq}
\ea
where $X_i\equiv\partial_i \ln a$. In the braneworld case, the long wavelength limit is advocated for consistently neglecting the projected Weyl tensor. This in turn is deeply intertwined with the other fundamental constraint, that is a standard continuity equation (\ref{Nconti}). In the presence of a source term surviving at all scales in the continuity equation, equation (\ref{conseq}) would not be valid and the quantity $\zeta_i$ would not be conserved. However, bulk physics affects only the small-scale or late-time cosmological structure and does not play a chief role during inflation. This result has been confirmed with several methods both analytically and numerically \cite{WMLL}, \cite{GoM}--\cite{koy03}. We shall come back to this issue in the final section\footnote{Also, the evolution equations for the SR parameters defined by equations (\ref{Npsepsi}) and (\ref{NHeta}), as well as all the expressions which will be derived through equations (\ref{evo1}) and (\ref{evo2}), depend on the continuity equation with no extra flux. However, one can define the SR parameters only in terms of the energy density with no reference to equation (\ref{Nconti}). Let ${\cal N}\equiv \int_{t_i}^t H(t')\rmd t'$ be the number of $e$-foldings counted forward in time, and a subscript ${\cal N}$ denote differentiation with respect to it. Then, for instance, the parameters $\epsilon(\rho)\equiv -q\rho_{{\cal N}}/(2\rho)$ and $\eta(\rho)\equiv -\rho_{{\cal NN}}/(2\rho_{\cal N})$ satisfy equation (\ref{evo1}) for an ordinary scalar field, and coincide with the usual SR quantities only when equation (\ref{Nconti}) holds.}.

For a scalar field, the curvature perturbation $\zeta$ on hypersurfaces with constant energy density coincides with the curvature perturbation ${\cal R}$ on comoving hypersurfaces, which as a vector quantity reads 
\be\label{Ri}
{\cal R}_i \equiv X_i-\frac{H}{\Pi}\partial_i\psi\,.
\ee
In the linear approximation, equation (\ref{Ri}) is the spatial gradient of
\be
{\cal R}_{\rm L}=-\Psi_4-\frac{H}{\Pi}\delta\psi\,,
\ee
where $\Psi_4 = -\delta a/a$ is the gauge-invariant potential perturbing the spatial part of the metric. Define
\be\label{zPi}
z\equiv\frac{a\Pi}{c_{\rm S} H^{\oteta}}= a \left(\frac{2\epsilon}{3qc_{\rm S}^2 H^\theta}\right)^{1/2},
\ee
where the speed of sound $c_{\rm S}$ is $c_{\rm S}=1$ for the ordinary scalar field and $c_{\rm S}=\sqrt{1-2\epsilon_T/(3q)}$ for the tachyon. The spatial gradient of the Mukhanov--Sasaki variable $u\equiv-z{\cal R}_{\rm L}$ corresponds to the linear limit of
\be
{\cal Q}_i \equiv -z\zeta_i = -z{\cal R}_i= \tilde{a}\left(\partial_i\psi-\frac{\Pi}{H}X_i\right), \label{Q}
\ee
where $\tilde{a}\equiv zH/\Pi$. For the ordinary scalar field, $\tilde{a}=a$. From equations (\ref{spgrad}), (\ref{zPi}) and (\ref{Q}) we get
\be\label{partz}
\partial_i z = -{\cal Q}_i+\tilde{a}\partial_i\psi [1+\oteta\epsilon-\eta+(\oteta-1)\varsigma^2]\,,
\ee
where the quantity
\be
\varsigma^2 \equiv \frac{q}{NH} \frac{\dot{c}_{\rm S}}{c_{\rm S}}= \frac{4q\epsilon_T\eta_T}{3q-2\epsilon_T}\,,
\ee
is second order in the SR parameters.


\section{Generalized Mukhanov equation and stochastic inflation}\label{nlmukeq}

The evolution equation for ${\cal Q}_i$ can be computed within the multi-field framework of \cite{RiS2}. Here we consider the same calculation for a single field $\psi$ in a patch given by equation (\ref{FRW}); the results will match each other in the 4D case. The first time derivative of equation (\ref{Q}) is
\be
\dot{\cal Q}_i=\frac{\dot{z}}{z}{\cal Q}_i=NH{\cal Q}_i \left[1+\oteta\epsilon-\eta+(\oteta-1)\varsigma^2\right].\label{dotQ}
\ee
Another derivation gives
\be\label{Qmuktemp}
\ddot{\cal Q}_i-\frac{\ddot{z}}{z}{\cal Q}_i=0\,.
\ee
However, it is more convenient to recast the equation of motion as
\be\label{Qeom}
\ddot{\cal Q}_i-F\dot{\cal Q}_i+\Omega{\cal Q}_i=0\,,
\ee
where
\ba
\fl F &\equiv& \frac{\dot{N}}{N}-NH\,,\\
\fl \Omega &\equiv& F\frac{\dot{z}}{z}-\frac{\ddot{z}}{z}=2+\left(3\oteta-1\right)\epsilon-3\eta-4\oteta\epsilon\eta+ \left(1+\oteta-\wteta\right)\oteta\epsilon^2+\eta^2+\xi^2+(\oteta-1)g_\varsigma\varsigma^2.\nonumber\\\fl
\ea
The extra tachyonic term is
\be
g_\varsigma \equiv 3+(\theta-1)\epsilon_T-2\eta_T+\left(\frac{\theta}{2}-1\right)\varsigma^2+\frac{(\varsigma^2)^.}{\varsigma^2}\,.
\ee
The expression for $\Omega$ is in accordance with the computation in the linear theory \cite{cal3} and, as that, is \emph{exact} in the SR parameters. The expression found in \cite{RiS2,RiS3} ($\theta=0$,$\,\oteta=1$) is recovered via the exact relation
\be
\frac{V''}{H^2}= 3(\epsilon_\phi+\eta_\phi)-\eta_\phi^2-\xi_\phi^2\,.
\ee
Equation (\ref{Qeom}) is equivalent to a generalized Mukhanov equation (with the Laplacian term dropped) when expressed via conformal time $\rmd\eta\equiv N\rmd t/a$; since $\rmd_t^2=(N/a)^2\rmd_\eta^2+(N/a)F\rmd_\eta$ and $\Omega=-(N/a)^2\rmd^2_\eta z/z$, one has
\be\label{Qmuk}
\left(\frac{\rmd^2}{\rmd\eta^2}-\frac{1}{z}\frac{\rmd^2z}{\rmd\eta^2}\right){\cal Q}_i=0\,.
\ee
In the linear approximation and in momentum space, equations (\ref{Qeom}) and (\ref{Qmuk}) hold for ${\cal Q}_i \approx \partial_i u\to \rmi k_iu_\mathbf{k}$, where a subscript $k$ denotes the Fourier mode with comoving wavenumber $k$. A standard  calculation in de Sitter spacetime gives the mean value of the quantum field $u_\mathbf{k}$:
\be\label{dSsol}
\sqrt{\langle|u_\mathbf{k}|^2\rangle}=\frac{aH}{\sqrt{2(kc_{\rm S})^3}}\,.
\ee

The equation of motion for $u_\mathbf{k}$ can be written as an equation for the coarse-grained part of $u_\mathbf{k}$ sourced via a stochastic noise term. The coarse-grained part of the Mukhanov variable is $u_\mathbf{k}^{(c)}=u_\mathbf{k}{\cal W}(kR)$, where ${\cal W}$ is the Fourier transform of a window function $W(|\mathbf{x}-\mathbf{x}'|/R)$ falling off at distances larger than $R$. The scale $R$ is of order of the comoving Hubble radius, $R=h(aH)^{-1}$, with $h>1$ so as to encompass the whole horizon. With $h$ sufficiently larger than 1 we can safely discard the $k^2$ term in the Mukhanov equation.

The final result is then extended to the non-linear gradient variable ${\cal Q}_i$ at large scales, yielding 
\ba
&& \ddot{\cal Q}_i-F\dot{\cal Q}_i+\Omega{\cal Q}_i=\xi_i(t,\mathbf{x})\,,\label{lange1}\\
&& \xi_i(t,\mathbf{x})=\int \frac{\rmd^3\mathbf{k}}{(2\pi)^{3/2}}\rmi k_i \rme^{\rmi\mathbf{k}\cdot\mathbf{x}}\left[\ddot{\cal W}+\dot{\cal W}(2\rmd_t-F)\right]u_\mathbf{k}\alpha(\mathbf{k})+{\rm c.c.}\,,\label{lange2}
\ea
where the superscript $(c)$ in ${\cal Q}_i$ is understood and $\alpha(\mathbf{k})$ is a complex stochastic quantity such that the ensemble average $\langle\alpha(\mathbf{k})\alpha^*(\mathbf{k}')\rangle= \delta^{(3)}(\mathbf{k}-\mathbf{k}')$; it simulates the continuous crossing of modes outside the horizon and their addition to the coarse-grained part. As they stand, equations (\ref{lange1}) and (\ref{lange2}) properly encode the full stochastic contribution.

Equation (\ref{lange1}) is the non-linear extension of the Langevin-type equation we used in the first-SR-order heuristic computation in synchronous gauge:
\be
\dot{\psi}^{(c)}(t)=-\frac{U'}{3H}-\frac{\ddot{\psi}^{(c)}(t)}{3Hc_{\rm S}^2}+\xi(\mathbf{x},t)\approx -\frac{U'}{3H}+\xi(\mathbf{x},t)\,,
\ee
where $U(\phi)=V(\phi)$ and $U(T)=\ln V(T)$. In this and the other equation one can see that there are basically two sources of non-linearity. The first one is the back-reaction of the field fluctuations on the background encoded in the noise term; the second one is the self-interaction of the scalar field represented by the potential contribution $\Omega$ (or $-U'/(3H)$). Therefore, \emph{a priori} the statistical distribution of ${\cal Q}_i^{(c)}$ ($\psi^{(c)}$) will be non-Gaussian, even if quantum fluctuations are random.


\section{Braneworld and tachyon bispectrum}\label{bispectra}

In order to compute the scalar spectrum and bispectrum, we fix the gauge to the time slicing with respect to which the $k$th mode crosses the horizon simultaneously for all spatial points \cite{SB2}. Then $t=\ln (aH)$, $NH=(1-\epsilon)^{-1}$, and $R=h\rme^{-t}$. In this gauge, the gradient curvature perturbation and $\partial_i z$ are, respectively,
\ba
{\cal Q}_i &=& \tilde{a}\partial_i\psi (1-\epsilon)\,,\\
\partial_i z &\approx& \left[\left(1+\oteta\right)\epsilon-\eta\right]{\cal Q}_i\\
             &=& -\case12 (n_{\rm s}-1){\cal Q}_i\,,\label{zgfix}
\ea
where in the last passage we have used the first-order patch SR expression for the scalar spectral index \cite{cal3,cal5}:
\be\label{noteta}
n_{\rm s}-1 = 2\eta-2\left(1+\oteta\right)\,\epsilon\,.
\ee
At first SR order,
\be
F \approx -1\,, \qquad \Omega \approx 2+3\left(\oteta+1\right)\epsilon-3\eta\,.
\ee
Equation (\ref{lange1}) can be expressed as a Langevin differential equation in the curvature perturbation:
\be
\ddot{\cal R}_i+\left(2\frac{\dot{z}}{z}-F\right)\dot{\cal R}_i=-\frac{1}{z}\,\xi_i(t,\mathbf{x})\,.
\ee
To lowest SR order and neglecting the $\ddot{\cal R}_i$ term,
\be\label{finR}
\dot{\cal R}_i \approx -\frac{1}{3z}\xi_i(t,\mathbf{x})\,,
\ee
where $c_{\rm S} \approx 1$ inside $\xi_i$ and $z$. The power spectrum is given by the solution of the linearized equation (\ref{finR}): at first order in a perturbative expansion,
\be
\dot{\cal R}_i^{(1)} \approx -\frac{\xi_i^{(1)}}{3z^{(0)}}\,,
\ee
where $z^{(0)}$ is $z$ defined on the homogeneous background. The time integration of $\xi/z$ from the initial time $t_i$ to $t$ is proportional to $B(kR)=\int_{t_i}^{t} \rmd t'(aH)^{-1}[\ddot{\cal W}+\dot{\cal W}(2\rmd_{t'}+1)](aH)=[1+(kR)^2/3]{\cal W}(kR)-[1+(kR_i)^2/3]{\cal W}(kR_i)$ to lowest SR order. Here we have used a Gaussian window function 
\be
{\cal W}=\exp(-k^2R^2/2)\,,\qquad \dot{\cal W}=(kR)^2{\cal W}\,,\qquad \ddot{\cal W}=[(kR)^2-2]\dot{\cal W}\,,
\ee
and the lowest-SR-order eigenvalue equation $\rmd_t u_\mathbf{k}=u_\mathbf{k}$. In the limit of asymptotic past ($t_i \to -\infty$) and future ($t\to +\infty$), $B(kR)\to 1$. The integration over $k$ yields ${\cal R}_{\rm L}\equiv{\cal R}^{(1)}=\partial^{-2}\partial^i {\cal R}_i^{(1)}$ and the lowest-order amplitude 
\be \label{Sdeg}
\langle|{\cal R}^{(1)}(k)|^2\rangle \approx \frac{3q}{2k^3}\frac{H^{2+\theta}}{2\epsilon}\,,
\ee
after a computation almost identical to that of \cite{RiS3}. At second order in the perturbation, 
$-3\dot{\cal R}_i^{(2)} =(z^{-1})^{(1)}\xi_i^{(1)}+\xi_i^{(2)}/z^{(0)}$. Since $\xi_i^{(2)}=\Or(\epsilon^2)$, the only surviving term at first SR order is, by the first-order version of equation (\ref{zgfix}),
\ba
\dot{\cal R}_i^{(2)} &\approx&  -(z^{-1})^{(1)}\frac{\xi_i^{(1)}}{3} =- \left(-\frac{\partial^{-2}\partial^j\partial_j z^{(1)}}{z^{(0)}}\right)\frac{\xi_i^{(1)}}{3z^{(0)}}\\
                     &=& -\case12 (n_{\rm s}-1){\cal R}^{(1)}\dot{\cal R}_i^{(1)}\,,\label{R2}
\ea
After computing the non-linear term ${\cal R}^{(2)}=-\case12 (n_{\rm s}-1)[{\cal R}^{(1)}]^2$, one is ready to write down the curvature perturbation at second order, 
\be
{\cal R} \approx {\cal R}^{(1)}+\case12{\cal R}^{(2)}\,,
\ee
and the bispectrum (\ref{fnl}). Actually, the integration of the window function inside the noise term corresponds to a non-linearity parameter with non-trivial momentum dependence, for which equation (\ref{nonlinear}) does not hold. When one of the momenta is negligible relative to the others, that is in the squeezed limit $k_3 \ll k_1,k_2$, by equations (\ref{R2}) and (\ref{fnl}) we get
\be
4f_{\rm NL} = (n_{\rm s}-1)+4f(\mathbf{k}_1,\mathbf{k}_2,\mathbf{k}_3) \approx n_{\rm s}-1\,.
\ee
Although we have not written explicitly the momentum structure $f(k)$, which at first SR order is the same as in \cite{RiS3}, we can draw some important conclusions.

\begin{enumerate}
\item[(i)] Tachyon and ordinary inflation generate the same non-Gaussian signature in the limit of collapsing momentum dependence and at first SR order. Outside this approximation the non-linearity parameter $f_{\rm NL}$ can depend on the type of scalar field and braneworld through the parameter $z$: $4f_{\rm NL}=(n_{\rm s}-1)+4f^{(\theta,\psi)}(\epsilon,\eta;\mathbf{k}_1,\mathbf{k}_2,\mathbf{k}_3)$. This is in agreement with the correspondence between the lowest-SR-order tachyon and ordinary observables.

\item[(ii)] Written in terms of $n_{\rm s}$, braneworld non-Gaussianity does not differ from the 4D picture, except perhaps in higher-order contributions. What changes is the inflationary model one has to impose in order to predict a given scalar spectral index. In this sense, we could regard equation (\ref{fns}) as a first-SR-order consistency equation joining the traditional set we explored elsewhere \cite{cal5}, since both $n_{\rm s}$ and $f_{\rm NL}$ (through the bispectrum) are observables \cite{gru04,CZ}\footnote{A small non-Gaussian component comes also from the three-point functions involving the graviton zero mode. Using the $z$ function for braneworld tensor perturbations $z\propto aH^{-\theta/2}$ \cite{cal5}, one finds a contribution proportional to the tensor amplitude and spectral index $n_{\rm t}=-(2+\theta)\epsilon$, in agreement with the 4D result \cite{mal03}. However, since the tensor amplitude is much smaller than the scalar one, we can neglect this term with respect to the scalar bispectrum.}.

\item[(iii)] On the other hand, one should note that the post-inflationary era greatly enhances non-Gaussianity, up to $f_{\rm NL}^{\rm post} =\Or(1)$ \cite{BMR1}--\cite{BMR3}\footnote{This is basically due to the fact that at second order the longitudinal gauge condition $\Phi_4^{(1)}-\Psi_4^{(1)}=0$ is modified as $\Phi_4^{(2)}-\Psi_4^{(2)}= 4\left(\Psi_4^{(1)}\right)^2$ at large scales, thus providing a non-trivial second-order correction to the Sachs-Wolfe effect \cite{ABMR}.} or even $f_{\rm NL}^{\rm post} \sim 50$ from a suitable preheating phase \cite{EJMMV}. As explained in \cite{BKMR}, the true observed non-linearity parameter is not the bare inflationary result (\ref{fns}). In addition to the post-inflationary contribution, one must consider angular averaging. The total observed $f_{\rm NL}$ is in fact, and at least, $f_{\rm NL}^{\rm obs}=\Or(1)+f_{\rm NL}$. Therefore the non-linear effect of braneworld or 4D inflation, if the SR approximation holds as we required, is always subdominant.
\end{enumerate}
Although these features might be obvious when inspecting patch cosmology, here we have derived them quantitatively\footnote{The authors of \cite{CZ} drew similar conclusions, claiming that equation (\ref{fns}) is a model-independent consistency equation under the assumption of single-field inflation and in the squeezed limit. However, see \cite{BKMR} (Sec. 8.4.1). Reference \cite{gru04} deals with the de Sitter case only.}. Moreover, the gradient+stochastic approach can be the basis for next-to-leading-order SR and perturbation calculations as well as for numerical simulations \cite{RiS2}.

According to the first-year WMAP analysis, the power spectrum fully characterizes the statistical properties of CMB anisotropies: that is, $f_{\rm NL}$ vanishes consistently. More precisely, the constraint on the non-linearity parameter is $-58<f_{\rm NL}<134$ \cite{kom03}, which does not discard not only inflationary non-Gaussianity, but also other models predicting a more robust effect, $f_{\rm NL} \gg 1$. See also \cite{GaW,cab04} for other analyses. The next-year WMAP data and the Planck satellite should significantly improve the accuracy of the measure, with the inclusion of polarization anisotropies: $f_{\rm NL}^{\rm min}({\rm WMAP})\sim 11-15$, $f_{\rm NL}^{\rm min}({\rm Planck})\sim 3-5$ \cite{BZ}.

We also note that in this class of models (single brane field and empty bulk) the contribution of $f(k)$ is below the present observational threshold. Actually, in order to experimentally detect and constrain the contribution of the momentum-dependent part of $f_{\rm NL}$, it should be much larger than $\Or(\epsilon)$, say $f(k)\sim \Or(10)$. Since the characteristic coefficient in $f(k)$ is $\Or(1)$ (see \cite{RiS3} for details), the exact form of this term is not relevant in such a set-up.


\section{Non-commutative bispectrum}\label{ncbisp}

Until now we have considered a commutative background throughout the whole spacetime. We can make a step further and phenomenologically assume that we have a 3-brane in which the stringy spacetime uncertainty relation (SSUR), $\Delta \tau\Delta x \geq l_{\rm s}^2\,,$ holds for all the braneworld coordinates $\{x^\nu\}$, $\nu = 0,1\,2\,3$, while the extra dimension $y$ along the bulk remains decoupled from the associated *-algebra. Here, $\tau \equiv\int\rmd t Na$,  ($\approx a/H$ in the SR regime), $x$ is a comoving spatial coordinate on the brane, and $l_{\rm s} \equiv M_{\rm s}^{-1}$ is the fundamental string scale. 

Non-commutative braneworld inflation arises when we impose a realization of the *-algebra on the brane coordinates. In order to diagonalize the non-commutative algebra and induce a pure 4D SSUR on the brane, one might fix the expectation values of the background fields of the fundamental theory such that the extra direction commutes, $[y,x^\nu]=0$. An algebra preserving the maximal symmetry of the FRW universe is $[\tau,x]=\rmi l_{\rm s}^2$ \cite{BH}, by which normal products in the action of the perturbation modes are replaced by *-products like
\begin{equation}\label{*prod}
(f*g)(x,\tau)=e^{-(\rmi l_{\rm s}^2/2)(\partial_x\partial_{\tau'}-\partial_{\tau}\partial_{x'})}f(x,\tau)g(x',\tau')\big|_{x'=x,\, \tau'=\tau}\,.
\end{equation}
The non-commutative algebra induces a cut-off $k_0$ roughly dividing the space of comoving wavenumbers into two regions, one encoding the ultra-violet (UV), small-scale perturbations generated inside the Hubble horizon ($H \ll M_{\rm s}$) and the other describing the infra-red (IR), large-scale perturbations created outside the horizon ($H \gg M_{\rm s}$). By definition, they correspond to the quasi-commutative and strongly non-commutative regime, respectively. 

The separate universe approach does not contrast with a non-commutative background. To understand this point, we can use the linear picture of \cite{WMLL}, and in particular their figure 1. The basic idea is that a comoving large-scale perturbation is independently specified in two comoving locally homogeneous regions separated by a distance $\lambda$, if the size $\lambda_{\rm s}\gtrsim H^{-1}$ of these regions is small with respect to $\lambda$. Perturbations are defined on a given background, that is a region of scale $\lambda_0$ much larger than our present horizon. Then the required hierarchy of scales is $\lambda_0\gg\lambda\gg\lambda_{\rm s}\gtrsim H^{-1}$. In the presence of a non-local algebra, the string scale can play the role of natural marker in the hierarchy. For instance, setting $\lambda_{\rm s}\sim l_{\rm s}$ we just consider the IR region of *-models. If $\lambda_{\rm s}\gtrsim H^{-1}>l_{\rm s}$, the previous argument is unchanged. 

Since the *-product (\ref{*prod}) does not involve homogeneous quantities (i.e., it preserves the FRW maximal symmetry), the Mukhanov equation for a non-commutative 4D \cite{BH} or braneworld \cite{cal4} scenario is, at linear order and large scales,
\be \label{muk}
\left(\frac{\rmd^2}{\rmd\teta^2}-\frac{1}{z_k}\frac{\rmd^2z_k}{\rmd\teta^2}\right)u_\mathbf{k}=0\,,
\ee
where the effective conformal time $\rmd \teta \equiv(a/a_\eff)^2\rmd\eta$, effective scale factor $a_\eff$, and measure $z_k$ are constructed through smeared, momentum-dependent versions of the scale factor $a_\pm\equiv a(\tau\pm kl_{\rm s}^2)$. In the separate universe approach, $a_\pm$ acquires a spatial dependence like the other quantities, $a_\pm(\tau)\to a_\pm(\mathbf{x},\tau)$.\footnote{Note that the correct procedure is first to smear the scale factor in a sufficiently small homogeneous neighbourhood, $a(\tau)\to a_\pm (\tau)$, and then to extend it to very large scales, $a_\pm(\tau)\to a_\pm(\mathbf{x},\tau)$. The `top-down' smearing $a(\mathbf{x},\tau)\to a_\pm(\mathbf{x},\tau)$ does not lead to equation (\ref{muk}).} The measure $z_k$ is given by the product of $z$ and a correction factor $f_z$ depending on the particular non-commutative model one is assuming \cite{cal4}. The de Sitter solution of equation (\ref{muk}) is the commutative solution (\ref{dSsol}) multiplied by $f_a^2\equiv (a_\eff/a)^2$, which is the relative rescaling of commutative to non-commutative conformal time.

The above discussion on conserved non-linear perturbations is not modified by the introduction of a fundamental length scale. Therefore we are tempted to directly generalize equation (\ref{muk}) with the gradient variable ${\cal Q}_i$.
Since the mass term $\rmd_\teta^2 z_k/z_k$ now depends on $k$, we define the Fourier mode of ${\cal Q}_i$ as
\be
{\cal Q}(\mathbf{k}) = \int \frac{\rmd^3 \mathbf{x}}{(2\pi)^{3/2}}\, \rme^{-\rmi \mathbf{k}\cdot\mathbf{x}}\partial^{-2}\partial^i{\cal Q}_i(\mathbf{x})\equiv -z_k{\cal R}(\mathbf{k})\,.
\ee
Then equation (\ref{muk}) holds for $u_\mathbf{k}\to {\cal Q}(\mathbf{k})$. One gets the $t$-version equation (\ref{Qeom}) by using the effective time $\rmd\tilde{t}=\rmd\teta a_\eff/N=\rmd t/f_a$ and setting $\tilde{N}=N$ without loss of generality:
\be
\rmd^2_{\tilde{t}}{\cal Q}(\mathbf{k})-\tilde{F}\rmd_{\tilde{t}}{\cal Q}(\mathbf{k})+\tilde{\Omega}{\cal Q}(\mathbf{k})=0\,,
\ee
where $F$ and $\Omega$ are
\ba
\tilde{F}     &=& f_aF-\dot{f}_a\,,\\
\tilde{\Omega} &=& \tilde{F}\frac{\rmd_{\tilde{t}}z_k}{z_k}-\frac{\rmd^2_{\tilde{t}}z_k}{z_k}=-\left(\frac{N}{a_\eff}\right)^2\frac{\rmd^2_\teta z_k}{z_k}=f_a^2(\Omega-\omega)\,,\\
\omega        &\equiv& \left[2\frac{\dot{z}}{z}\left(\frac{\dot{f}_a}{f_a}+\frac{\dot{f}_z}{f_z}\right)-F\frac{\dot{f}_z}{f_z}\right]+ \left(\frac{\ddot{f}_z}{f_z}+2\frac{\dot{f}_a}{f_a}\frac{\dot{f}_z}{f_z}\right)\,.\label{ncomeg}
\ea
In the commutative limit, $\tilde{F}\to F$, $\omega\to 0$, and $\tilde{\Omega}\to\Omega$. The term in square brackets in equation (\ref{ncomeg}) contains the $\Or(\epsilon)$ contribution of $\tilde{\Omega}$, since $\dot{f}_a/f_a=\Or(\epsilon)=\dot{f}_z/f_z$. In the IR limit (strongly non-commutative regime), $\omega$ and its components loose their momentum dependence \cite{cal4}. The Langevin equation for the curvature perturbation reads
\be\label{langRnc}
\rmd^2_{\tilde{t}}{\cal R}(\mathbf{k})+\left(2f_a\frac{\dot{z}_k}{z_k}-\tilde{F}\right)\rmd_{\tilde{t}}{\cal R}(\mathbf{k})=-\frac{1}{z_k}\,\xi(\tilde{t},\mathbf{k})\,.
\ee
The non-commutative version of equation (\ref{partz}) gets an extra term from the redefinition of $z$; in momentum space and at first SR order, $z_k \approx -{\cal Q}(\mathbf{k})+\tilde{a}_\mathbf{k}\psi_\mathbf{k}[1+\oteta\epsilon-\eta+\dot{f}_z/(NHf_z)]$. 

With the gauge choice $\teta\approx -(aHf_a^2)^{-1}=-\rme^{-\tilde{t}}$, by definition one has $\partial_i \tilde{t}=0$ on surfaces of constant non-commutative time. Then 
\be
NHf_a=\frac{1-2\dot{f}_a}{1-\epsilon}= 1+\Or(\epsilon)\,,\qquad \tilde{F}=-1+\Or(\epsilon^2)\,,\qquad \rmd_{\tilde{t}}u_\mathbf{k}=u_\mathbf{k}\,.
\ee
Neglecting the second-derivative term in equation (\ref{langRnc}), one has
\be
\rmd_{\tilde{t}}{\cal R}(\mathbf{k}) \approx -\frac{1}{3z_k}\xi(\tilde{t},\mathbf{k})\,.
\ee
With the procedure of the last section, at lowest SR order and first perturbative order, one obtains the non-commutative power spectrum found in \cite{BH,cal4}, which is the commutative amplitude (\ref{Sdeg}) corrected by a factor $\Sigma^2\equiv (f_a^2/f_z)^2$. Actually, if one wanted to go to coordinate space, the integration over momenta should be performed up to the UV cut-off $k_0$, while the characteristic scale $R$ should be pushed towards the asymptotic limit $R \sim k_0^{-1}$ at the infinite future. However, the approximation $k_0\to \infty$ fits well at this stage of accuracy.

The gauge-fixed variables ${\cal Q}$ and $z_k$ are
\ba
{\cal Q}(\mathbf{k}) = \tilde{a}_\mathbf{k}\psi_\mathbf{k} (1-\epsilon+2\dot{f}_a)\,,\\
z_k \approx \left[\left(1+\oteta\right)\epsilon-\eta+f_a\frac{\dot{f}_z}{f_z}-2\dot{f}_a\right]{\cal Q}(\mathbf{k})= -\frac12 (n_{\rm s}-1){\cal Q}(\mathbf{k}).\qquad\label{zkgfx}
\ea
At first SR order, the spectral index (\ref{noteta}) has acquired an extra term
\be\label{signga}
\sigma\epsilon\equiv\frac{\rmd\ln\Sigma^2}{\rmd\ln k}=\frac{2}{NH}\left(2\frac{\dot{f}_a}{f_a}-\frac{\dot{f}_z}{f_z}\right)\approx 2f_a\left(2\frac{\dot{f}_a}{f_a}-\frac{\dot{f}_z}{f_z}\right),
\ee
where $\sigma$ is constant in the IR limit. A computation of the second-order curvature perturbation, by equation (\ref{zkgfx}), gives again equation (\ref{fns}), with the spectral index now depending on the non-commutative parameter $\sigma$.


\section{Conclusions}\label{concl}

In the context of tachyon and braneworld scenarios, early-Universe large-scale perturbations can leave an imprint different to that of standard four-dimensional inflation in the microwave sky. Under certain assumptions, we have shown that the linear cosmological spectrum comes from the first term of a gradient perturbative expansion of a non-linear curvature perturbation satisfying a generalized Mukhanov equation of motion. The bispectrum of this quantity, which involves it at second order, governs the non-Gaussian signature eventually detectable in the CMB. Neglecting the projected Weyl tensor on the brane, we have found that the non-linearity parameter $f_{\rm NL}$ is proportional to the scalar spectral index and therefore unobservable, in agreement with past 4D calculations and even in the case of a particular realization of a non-commutative spacetime. 

A way to generate a stronger non-Gaussian signal in the braneworld context might be to include the Weyl contribution (both in the Friedmann and continuity equations), but as we have said we should expect it to play a negligible role in the long wavelength limit. Nevertheless, this issue will deserve further attention for at least two good reasons. The first is that Weyl damping was considered and shown only in the case of linear perturbations, while the stochastic Langevin equation for the curvature invariant holds at all orders. The second is that bulk physics intrinsically provides a noise source to the Mukhanov equation through an infinite tower of Kaluza--Klein scalar modes dominating at short wavelengths \cite{KLMW}. Therefore, while equation (\ref{Qeom}) would keep on being valid, an important contribution to the stochastic noise term (\ref{lange2}) might be lacking in the present analysis. In this case the consistency equations, included equation (\ref{fns}) even in the squeezed limit, would be spoiled anyway \cite{der04}.

Therefore one should expect to find the most interesting non-Gaussian effects, if any, in the spectral regions we have neglected so far, namely, around the time of horizon-crossing or after, at sub-horizon scales. Here one should note a caveat hidden in the above approach. The stochastic noise given by the random field $\alpha(\mathbf{k})$ mimics at-horizon and sub-horizon physics without a more fundamental knowledge of it, so that this method implicitly misses cubic interactions around horizon crossing. As stressed in
\cite{RiS4}, this might be the source of the discrepancy between \cite{RiS3} and \cite{mal03} as regards the shape of $f(k)$. In the latter paper this physics is under control and the resulting answer perhaps relies upon more solid ground. For these reasons, one might be concerned with whether the formalism we chose is not the
most suitable to investigate signatures of braneworld non-Gaussianity. However, it is probably one of the most calculationally accessible methods at present time, which gives the correct momentum-independent result anyway. In a more complete treatment of braneworld non-Gaussianity, the $k$-dependent part of the non-linear parameter might change relative to the 4D result and should be presented explicitly, at least for estimating its magnitude. As stated in section \ref{bispectra}, only when $f(k)\gtrsim 10$ could this term be important from the observational point of view. Only if this were the case in the stochastic approach would the search for an alternative method be worth the extra effort.

To conclude, we would like to be more tentative in guessing the magnitude of $f_{\rm NL}$ when the bulk physics is taken into account. In general, the effect is rather model dependent and there is still no univocal result.
For example, large non-Gaussianities can be obtained in the `D-cceleration' scenario \cite{AST} or for specific choices of the Weyl tensor in the RS brane \cite{der04}. However, in \cite{KMW} the Weyl noise term in the Mukhanov equation for the RS brane has been calculated and is $\Or(\epsilon)$ (see their equations (103)--(106)), which might imply a small extra contribution to the three-point correlation function, not larger than the spectral index. This issue should be checked and is now under investigation.


\ack

The author thanks Claudia de Rham, Eiichiro Komatsu and Filippo Vernizzi for useful discussions.


\end{document}